\documentclass{article}

\input{tcilatex}
\begin{document}

\title{Comment on \textquotedblleft Wave-particle duality revisited: Neither
wave nor particle\textquotedblright\ (arXiv:1907.09836v1 [quant-ph])}
\author{Luiz Carlos Ryff \\
\textit{Instituto de F\'{\i}sica, Universidade Federal do Rio de Janeiro,}\\
\textit{Caixa Postal 68528, 21041-972 Rio de Janeiro, Brazil}\\
E-mail: ryff@if.ufrj.br}
\maketitle

\begin{abstract}
In a recent article Jan Sperling, Syamsundar De, Thomas Nitsche, Johannes
Tiedau, Sonja Barkhofen, Benjamin Brecht, and Christine Silberhorn\ discuss
the wave-particle duality\ using an experiment to demonstrate that
\textquotedblleft neither the wave nor the particle description is
sufficient to predict the outcomes of quantum-optical
experiments.\textquotedblright\ I would like to draw your attention to two
previous papers that discuss feasible experiments in which a photon does not
seem to behave either as particle or as wave and make some brief
considerations on the topic.

\end{abstract}

In a recent article entitled \textquotedblleft Wave-particle duality
revisited: Neither wave nor particle\textquotedblright\ \textrm{[1] }Jan
Sperling, Syamsundar De, Thomas Nitsche, Johannes Tiedau, Sonja Barkhofen,
Benjamin Brecht, and Christine Silberhorn argue that \textquotedblleft
neither the wave nor the particle description is sufficient to predict the
outcomes of quantum-optical experiments.\textquotedblright\ According to
them, \textquotedblleft Using squeezed light, it is then confirmed that
measured correlations are incompatible with either picture. Thus, within one
single experiment, it is proven that neither a wave nor a particle model
explains the observed phenomena.\textquotedblright\ They also emphasize that
\textquotedblleft Maybe to avoid giving up classical concepts, the popular
interpretation still is that, depending on the context, quantum systems can
be described in a particle or wave model; we aim at disproving this
belief.\textquotedblright\ I would like to draw your attention to two
previous papers entitled \textquotedblleft Neither particle-like nor
wave-like behaviour of a photon\textquotedblright\ \textrm{[2]} and
\textquotedblleft Two-photon interference without intrinsic
indistinguishability\textquotedblright\ \textrm{[3] }that discuss feasible
experiments in which a photon seems to behave neither as particle nor as
wave. Differing from the argument presented by the above authors, that uses
squeezed states, ours follows from what seems to be a conflict between the
wave-like and nonlocal properties of entangled photons. Initially, it is
important to make it clear that the mathematical formalism of quantum
mechanics provides the correct results for the experiments being
considered.\ The \textquotedblleft wave-or-particle\textquotedblright\
picture only serves to come up with a story a posteriori. Following
Heisenberg, we have coexisting\ \textquotedblleft
potentialities.\textquotedblright\ How these potentialities are
\textquotedblleft actualized\textquotedblright\ (so to speak) is still
considered a mystery, which constitutes the quantum measurement conundrum,
despite some divergent -- and also diverging from each other -- opinions. As
shown in ref. \textrm{[2]} and \textrm{[3]}, in some situations trying to
trace back the history of the photon leads to a contradiction, since
different and \textit{mutually exclusive} interpretations are possible 
\textrm{[4]}.\textrm{\ }\ We may try to circumvent this difficulty by
introducing a preferred frame, as suggested by Bell and Bohm \textrm{[5]},
and assuming that a photon always propagates as a wave and is detected as a
particle \textrm{[6]}.


\begin{thebibliography}{9}
\bibitem{6} It might also be interesting to examine the possibility of
describing the propagation of the photon using a nonlocal pilot wave
interpretation.
\end{thebibliography}
\end{document}